\documentclass[twocolumn]{revtex4}
\usepackage{graphicx}

\begin{document}
\pagestyle{empty} 
\title{Role of surface roughness in superlubricity}

\author{U. Tartaglino}
\affiliation{IFF, FZ-J\"ulich, 52425 J\"ulich, Germany}
\affiliation{Democritos National Simulation Center, Via Beirut 2, 34014 Trieste, Italy}

\author{V.N. Samoilov}
\affiliation{IFF, FZ-J\"ulich, 52425 J\"ulich, Germany}
\affiliation{Physics Faculty, Moscow State University, 117234 Moscow, Russia}

\author{B.N.J. Persson}
\affiliation{IFF, FZ-J\"ulich, 52425 J\"ulich, Germany}

\begin{abstract}

We study the sliding of elastic solids in adhesive contact with flat and
rough interfaces. We consider the dependence of the
sliding friction on the elastic modulus of the solids.
For elastically hard solids with planar surfaces with incommensurate surface
structures we observe extremely low friction (superlubricity), which
very abruptly increases as the elastic modulus decreases. We show that
even a relatively small surface roughness may completely kill the superlubricity state.  

\end{abstract}
\maketitle

%%%%%%%%%%%%%% main text %%%%%%%%%%%%%%%%

%\vskip 0.5cm

%{\bf 1. Introduction}
\section{Introduction}
\label{sec1}

Friction between solid surfaces is a common phenomenon
in nature and of extreme importance 
in biology and technology\cite{P0}. At the most fundamental level friction
is (almost) always due to elastic instabilities at the sliding interface.
At low sliding velocity an elastic instability first involves (slow) elastic loading,
followed by a rapid rearrangement, where the speed of the rearrangement is much faster than, and
unrelated to, the loading (or sliding) velocity. During the fast rearrangement the 
elastic energy gained during the loading phase
is converted into irregular heat motion. 
The exact way of how the energy is ``dissipated'' has
usually a negligible influence on the sliding friction force, 
assuming that the dissipation occurs so fast that
no memory of it remains during the next elastic loading event. 
There are many possible origins of elastic instabilities, e.g., it may involve
individual molecules or, more likely, group of molecules or 
``patches'' at the interface which we have denoted by stress domains\cite{P11,P2,Caroli,Caroli1}.
The most
fundamental problem in sliding friction is to understand the physical origin 
and nature of the elastic instabilities.  

\begin{figure}
  \includegraphics[width=0.40\textwidth]{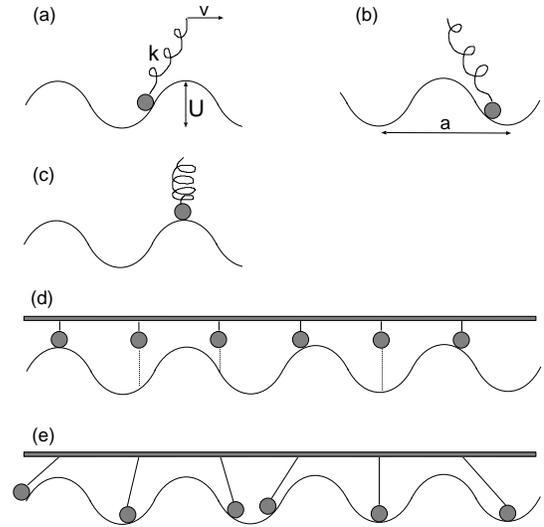} %{instabil.ps}
  \caption{ \label{instabil}
(a) A particle moving in a corrugated potential. The particle is connected to a
spring which is pulled with the velocity $v$.
(b) Stick-slip motion with a soft spring, $ka^2<U$.
(c) Continuous sliding with a stiff spring, $ka^2>U$.
(d) Elastically stiff solid sliding on a rigid corrugated substrate.
(e) Elastically soft solid sliding on a rigid corrugated substrate: here
the atoms can rearrange to occupy binding positions.
}
\end{figure}

Elastic instabilities occur only if the lateral corrugation of the 
interaction potential between the solid walls is high enough,
or the elastic modulus of the solids small enough. 
Roughly speaking, elastic instabilities can only occur if a characteristic elastic energy is 
smaller than a characteristic binding (or rather barrier height) energy.
To understand this, consider
the simple model illustrated in Fig.~\ref{instabil}.
In (a) a particle or atom is moving in a corrugated (substrate) potential.
Connected to the particle there is a spring (spring constant $k$)
which is pulled with the velocity $v$. If the spring is
soft enough, or the potential barrier height $U$ is high enough, i.e., $U > ka^2$, the particle will perform 
stick-slip motion [Fig.~\ref{instabil}(a),(b)], 
involving slow elastic loading followed by rapid slip and dissipation
of the (elastic) spring energy. In this case the (time averaged) force on the particle is independent of $v$.
However, in the opposite case $U < ka^2$ 
[Fig.~\ref{instabil}(c)], 
the particle will 
follow 
the drive with a velocity which is always comparable to $v$. In particular,
when the drive is on-top of the barrier so will the particle be
[Fig.~\ref{instabil}(c)].
In this case no rapid motion
will occur and the (time averaged) friction force acting on the particle is proportional to $v$.

In a more realistic situation one must consider the whole interface. In this case, depending on the
elasticity and lateral barriers and the size of the contact area, 
elastic instabilities may or may not occur\cite{Aubry}.
Assume first that 
an elastically very stiff solid slides on a rigid corrugated substrate,
[Fig.~\ref{instabil}(d)]. 
In this case the atoms at the
bottom surface cannot adjust to the corrugated substrate potential, and (for an incommensurate system)
as some atoms move downhill other atoms 
move uphill in such a way that the total energy is constant. 
Thus, no elastic instabilities will occur during sliding, resulting in a very low sliding friction;
this state has been termed {\it superlubric}\cite{Japan}.
However, when the block is elastically soft
[Fig.~\ref{instabil}(e)], 
the atoms can rearrange themselves so that at any moment in time almost all
the atoms occupy positions close to the minima of the substrate potential. 
During sliding rapid jumps will
occur from time to time where a particle changes potential well. In this case the friction 
is high and (at zero temperature) remains finite
as the sliding velocity $v\rightarrow 0$.         

It is well known that elastically hard solids tend to exhibit smaller sliding friction
than (elastically) soft materials\cite{Elisa}. One extreme example is diamond which under normal
circumstances exhibits very low kinetic friction coefficient, of the order of 0.01, 
when diamond is sliding on diamond.
This can be explained by the nearly absence of elastic instabilities because of the 
elastic hardness of the material. However, if clean diamond is sliding on clean diamond in ultrahigh
vacuum, a huge friction (friction coefficient of the order of $\mu \approx 10$) 
is observed\cite{Flipse}. 
The reason is that the clean surfaces have dangling bonds
(which are passivated by hydrogen and oxygen in the normal atmosphere) so that the interaction
between the two diamond surfaces is very strong and elastic instabilities (and wear processes)
can occur resulting in a very large friction.

\begin{figure}
  \includegraphics[width=0.40\textwidth]{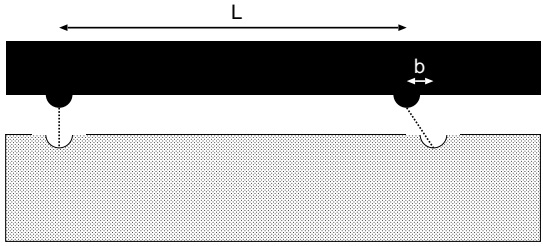} %{elast1D.ps}
  \caption{ \label{elast1D}
An elastic bar with two binding sites (bump). The corresponding binding
cavities of the substrate do not match exactly the positions of the
binding sites, causing a competition between the energy cost to stretch
the elastic bar and the binding energy.
}
\end{figure}

It is important to note that 
even if solids are too stiff for elastic instabilities to occur on short length scale, 
the ratio between the effective elasticity 
and the amplitude of the lateral corrugation of the binding potential 
may decrease when the system is studied at a longer length scale, which may
make elastic instabilities possible on a longer length scale\cite{SSC,Nozier}.
To illustrate this, in Fig.~\ref{elast1D} we show a one-dimensional (1D) case, where
an elastic bar (cross section area $A$) with two binding sites (bumps) is in
contact with a substrate with two binding sites (cavities). When the binding
sites on the elastic bar overlap with the binding sites (cavities) on the (rigid) substrate, 
the binding energy
$U$ is gained. In order to gain this binding energy the segment of the elastic bar between
the bumps (length $L$) must elongate by the distance $b$. Thus the strain in the segment is
$b/L$ and the elastic energy stored in the elongated segment is $U_{\rm el} = V E (b/L)^2/2$
where the volume $V=LA$. Thus, 
$U_{\rm el} = A E b^2/(2L)$ which decreases as the length of the segment $L$ increases. 
It follows that only when $L > AEb^2/(2U)$ will the bound state have 
a lower energy than the non-bound state. Thus, only on a large enough length scale will the solid be
elastically soft enough for elastic instabilities to occur. 
In most practical cases one is not interested in a 1D situation but rather in
semi-infinite solids, which are intermediate between the 2D and 3D case. 
For surfaces with randomly distributed binding centers
this situation is much more complex than for the 1D case
because the effective elasticity changes as quickly with the
lateral length scale as does the effective amplitude 
of the lateral corrugation of the binding potential (which from random walk arguments\cite{P0}
scales as $L$)\cite{SSC,Nozier}. A detailed analysis of this situation indicates, however, that if no
elastic instability can occur at short length scale it is very unlikely that elastic
instabilities will occur on any length scale of 
practical importance, except perhaps in the context of earthquakes
\cite{SSC,Nozier}. 
If instead of randomly distributed binding sites one assumes incommensurate
surfaces, one would expect even weaker pinning effects, and it can be argued that
in this case the ratio between the effective elasticity
and the amplitude of the lateral corrugation of the binding potential {\it increases} as $\sim L$
so that if no elastic instabilities occur at short length scale they cannot occur
at any length scale\cite{Muser}. 
Below we will present numerical results where elastic instabilities do occur
also for (nearly) incommensurate structures, but in these cases one of the solids is elastically
very soft so that instabilities can occur on a short length scale. 

The discussion above has focused on clean surfaces and zero temperature. Temperature is unlikely to
have any drastic influence on superlubricity. However, it may have a strong influence on
sliding dynamics when elastic instabilities occur. As soon as $T > 0 \ {\rm K}$, thermal noise is
able to activate jumps over the barrier, i.e., to provoke {\it premature jumps} before the 
(zero temperature) instability point is reached. It has been shown experimentally\cite{Caroli1,Meyer1} and
theoretically\cite{P2,Activated} that this has a crucial influence on friction dynamics 
at low sliding velocity. 
Similarly, weakly bound adsorbed atoms and molecules have a large influence on the 
sliding dynamics,  and may strongly
increase the friction force\cite{he1999} as the mobile adsorbates can adjust themselves in the corrugated
potential between the block {\it and} the substrate, giving rise to strong pinning effects.
In this paper we will not address the role of adsorbates or non-zero temperature, but we will focus on the
simplest case of clean surfaces at zero temperature. 

Recently, superlubricity has been observed during sliding of graphite on
graphite: in the
experiment described in Ref.~\cite{Frenken} a tungsten tip with a graphite
flake attached to it is slid on an atomically flat
graphite surface. When the flake is in registry with the substrate stick-slip motion and large
friction are observed. When the flake is rotated out of registry, the forces felt by the different atoms
start to cancel each other out, causing the friction force to nearly vanish, and the contact to
become superlubric.

Graphite and many other layered materials are excellent dry lubricants. The most likely reason for this is that
the solid walls of the sliding objects get coated by graphite flakes or layers with different orientation
so a large fraction of the graphite-graphite contacts will be in the superlubric state. This will 
lead to a strong reduction in the average friction. However, the coated solid walls are unlikely to
be perfectly flat and it is important to address how surface roughness may influence the superlubric state.
In this paper we will show that even a relatively small surface roughness may kill the superlubric state.

Lubrication by graphite flakes may even occur for diamond-like carbon (DLC) 
coatings, which may exhibit very
low friction. Indeed Liu et al\cite{Liu} have observed that a graphitized tribolayer is formed on top
of diamond-like carbon coatings. Thus, also the excellent lubrication properties of
DLC films might be caused by superlubric graphite contacts. We also note that
DLC films are very hard and this
will reduce the chance for elastic instabilities to occur\cite{Erdemir}.

In this paper we present atomistic Molecular Dynamics calculations of the sliding
dynamics for contacting 
elastic solids with (nearly) incommensurate surface lattice structures.
We consider both flat and rough surfaces.
We consider the dependence of the
sliding friction on the elastic modulus of the solids.
For elastically hard solids with flat surfaces and incommensurate surface
structures we observe extremely low friction (superlubricity), which
very abruptly increases as the elastic modulus is diminished. We show that
even a small surface roughness may completely kill the superlubric state.  
In order to study large systems we use a recently developed 
multiscale approach\cite{Yang}
to contact mechanics where the number of dynamical variables scales like $\sim N^2$
rather than as $\sim N^3$, where $N\times N$ is the number of atoms in the nominal
contact area. 

%\vskip 0.5cm
%{\bf 2. Multiscale Molecular Dynamics}
\section{Multiscale Molecular Dynamics}
\label{sec2}

Let us discuss the minimum block size necessary in a computer simulation
for an accurate description of the
contact mechanics between two semi-infinite elastic solids with nominally flat surfaces. 
Assume that the surface roughness power spectrum has a roll-off
wavevector $q=q_0$ corresponding to the roll-off wavelength $\lambda_0 = 2 \pi /q_0$. 
In this case the minimum block must extend $L_x \approx \lambda_0$
and $L_y \approx \lambda_0$ along the $x$ and $y$ directions.
Furthermore, the block must extend at least a distance $L_z \approx \lambda_0$ 
in the direction 
perpendicular to the nominal contact area, since the surface roughness with
wavelength $\lambda_0$ affects the elastic block up to such a distance.
Thus, the minimum block is a cube with the side $L=\lambda_0$.

\begin{figure}
  \includegraphics[width=0.4\textwidth]{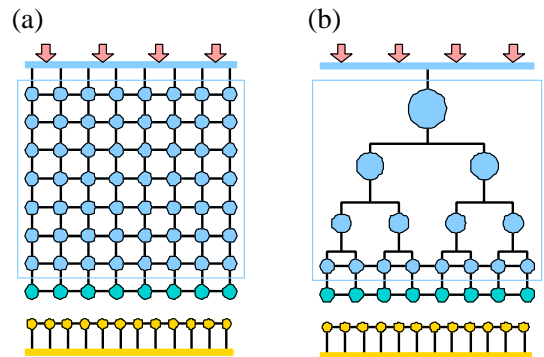} %{sb_nosb.eps}
  \caption{\label{smartblock}
          Schematic structure of the model. (a) The fully atomistic model.
          (b) The multiscale {\it smartblock} model, 
          where 
          the solid in (a) is coarse grained 
          by replacing groups of atoms with bigger ``atoms''.
          }
\end{figure}

As an example, if $\lambda_0$ corresponds to 1000 atomic spacings,
one must at least consider a block with $1000\times 1000$ atoms
within the $xy$ contact plane, i.e., one would need to study the elastic deformations
in a cubic block with at least $10^9$ atoms.
However, it is possible to drastically reduce the number of 
dynamical variables without loss of accuracy if one notes that an interfacial
roughness with wavelength $\lambda$ will give rise to a deformation field in the block which
extends a distance $\lambda$ into the solid, and which does not have any significant variation
over distances much smaller than $\lambda$.
Thus when we study the deformation a distance $z$ into the block
we do not need to describe the solid on the atomistic level, but we can coarse-grain 
the solid by replacing groups of atoms with bigger ``atoms'' as indicated schematically
in Fig.~\ref{smartblock}. 
If there are $N\times N$ atoms in the nominal contact area
one need $n\approx {\rm ln} N$ ``atomic'' layers in the $z$-direction.
Moreover the number of atoms in each layer decreases in a geometric progression
every time the coarse graining procedure is applied, so that the total number of
particles is of the order of $N^2$ instead of $N^3$. This results in a huge reduction in the
computational time for large systems.
This multiscale approach may be implemented in various ways, 
and in Ref.~\cite{Yang} we outline the procedure
we have used in this paper
which we refer to as the  {\it{smartblock}}.

The model presented above should accurately describe the deformations in the
solids as long as the deformations vary slowly enough with time.
However the phonons with short wavelength cannot propagate in the coarse
grained region because the model does not implement enough degrees of
freedom for them. The short wavelength phonons get scattered back when
they reach the coarse grained region, and this cannot be avoided within
a standard molecular dynamics approach. However this is not a serious
limitation for the present work: the static friction and the onset of
sliding are still unaffected; moreover the superlubric sliding at low
speed does not dissipate energy into phonons.
On the other hand, if one had to study the spectrum of dissipated
phonons, more advanced techniques to adsorb the energy of the phonons
without back scattering have to be considered\cite{cai,E}. 
The only dissipation mechanism that we employed in our simulations is a
Langevin thermostat at $T=0$ K (i.e., a viscous friction term) acting
only on the atoms of the block far from the contact region.

Figure \ref{smartblock} illustrates a case where the block is in the form of a cube
with atomically flat surfaces. It is possible to obtain curved surfaces of
nearly arbitrary shape by ``gluing''
the upper surface of the block to a hard curved surface profile. This was described in detail
in Ref.~\cite{Persson_JCP2001}. The elastic modulus and the shear modulus of the solid can be fixed at
any value by proper choice of the elongation and bending spring constants for the springs
connecting the atoms
(see Refs.~\cite{Persson_JCP2001} and \cite{Yang}). 
% The upper surface of the smartblock can be moved with arbitrary
% velocity in any direction, or an external force of arbitrary magnitude can be applied
% to the upper surface of the smartblock. We have also studied sliding friction problems where
% the upper surface of the smartblock is connected to a spring which is pulled in some prescribed way.
% The computer code also allows for various lubricant fluids between the solid 
% surfaces of the block and the substrate.
% Thus the present model is extremely flexible and can be used to study many interesting
% adhesion and friction phenomena, which we will report on elsewhere.

We note that with respect to contact mechanics,
when the slopes of the surfaces are small, i.e. when the surfaces are 
almost horizontal, one of the two surfaces
can be considered flat, while the profile of the other surface has to be 
replaced by the difference of the two original profiles\cite{Jon}. Thus, if the
substrate has the profile $z=h_1({\bf x})$ and the block has the profile
$z=h_2({\bf x})$, then we can replace the actual system with a fictive
one where the block has an atomically smooth surface while the substrate
profile $ h({\bf x}) = h_2({\bf x})-h_1({\bf x})$. Furthermore, if the original
solids have the elastic modulus $E_1$ and $E_2$, and the Poisson ratio $\nu_1$ and
$\nu_2$, then the substrate in the fictive system can be treated as
rigid and the block as elastic with the elastic modulus $E$ and Poisson
ratio $\nu$ chosen so that $(1-\nu^2)/E = (1-\nu_1^2)/E_1+(1-\nu_2^2)/E_2$.

The results presented below have been obtained for an elastic flat block sliding on
a rigid substrate. We considered both flat and rough substrates.
The atoms in the bottom layer of the block form a simple square lattice
with lattice constant $a$. The lateral dimensions $L_x=N_xa$ and $L_y=N_ya$. 
For the block, $N_x=N_y=48$. Periodic boundary conditions are
applied in the $xy$ plane.  
The lateral size of the block is equal to that of the substrate,
but for the latter we use different lattice constant $b \approx a/\phi$, where $\phi=(1+\sqrt{5})/2$
is the golden mean, in order to avoid the formation of commensurate
structures at the interface. For the substrate, $N_x=N_y=78$.
The mass of a block atom is 197 a.m.u.\ and
the lattice spacing of the block is $a=2.6~\mbox{\AA}$, so to get the same
atomic mass and density of gold. The lattice spacing of the substrate 
is $b=1.6~\mbox{\AA}$. 
We consider solid blocks with different Young's moduli 
from $E=0.2 \ {\rm GPa}$ up to $1000 \ {\rm GPa}$. The Poisson
ratio used for the block is $\nu = 0.3$. 

The atoms at the interface between the block and the substrate interact with 
the potential
\begin{equation}
 \label{potential}
 U(r)=4 \epsilon \left [ \left ({r_0\over r}\right )^{12}-\alpha 
 \left ({r_0 \over r}\right )^{6} \right],
\end{equation}
where $r$ is the distance between a pair of atoms.
When $\alpha = 1$, Eq.~(\ref{potential}) is the standard
Lennard-Jones potential. The parameter $\epsilon$ is the binding energy
between two atoms at the separation $r=2^{1/6} r_0$.
When we study contact mechanics and friction without adhesion 
we put $\alpha = 0$. In the calculations presented below we have used $r_0= 3.28~\mbox{\AA}$
and $\epsilon = 40 \ {\rm meV}$, which (when $\alpha = 1$) gives an 
interfacial binding energy (per unit area)\cite{Is}
$\Delta \gamma \approx 4\epsilon/a^2 \approx 23.7 \ {\rm meV/\mbox{\AA}^2}$. 

\begin{figure}
  \includegraphics[width=0.50\textwidth]{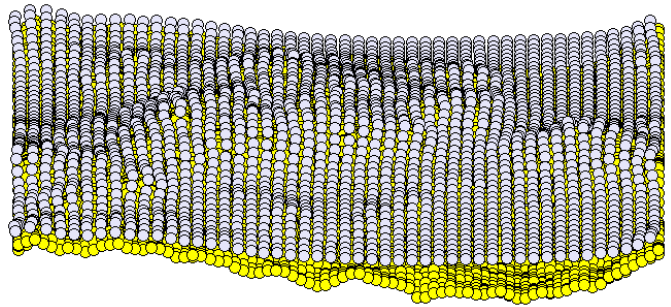} %{contact_view_s11b_Fractal_inp_dat00_cut.eps}
  \caption{ \label{contact1}
The contact between an elastic block with a flat surface and a rough 
{\it rigid} substrate. Only the interfacial layers of atoms are shown.
The elastic modulus of the block is $E=100 \ {\rm GPa}$. The substrate
is self-affine fractal with the root-mean-square roughness $3~\mbox{\AA}$,
fractal dimension $D_{\rm f} = 2.2$ and roll-off wavevector
$q_0=3q_L$, where $q_L= 2\pi /L_x$.
The substrate and block interfacial atomic layers consisted of 
$78\times 78$ and $48\times 48$ 
atoms, respectively. The applied pressure $p=10 \ {\rm 
GPa}$. Note the elastic deformation of the block, and that the real contact 
area is smaller than the nominal contact area. 
}
\end{figure}

As an illustration, in Fig.~\ref{contact1} we show
the contact between a flat elastic block (top) 
and a randomly rough 
{\it rigid} substrate (bottom). Only the interfacial block and substrate
atoms are shown. The substrate
is self-affine fractal with the root-mean-square roughness $3~\mbox{\AA}$
(see Sec.~\ref{sec3}). 
Note the elastic deformation of the block, and that non-contact regions
occur in the ``deep'' valleys of the substrate. Actually the real contact 
area is smaller than the nominal contact area. 

%\vskip 0.5cm
%{\bf 3. Self-affine fractal surfaces}
\section{Self-affine fractal surfaces}
\label{sec3}

Consider a solid with a nominally flat surface. Let $x,y,z$ be a coordinate system with the
$x,y$ plane parallel to the surface plane.
Assume that $z=h({\bf x})$ describes the surface height profile, where ${\bf x} = (x,y)$
is the position vector within the surface plane. The most important property characterizing 
a randomly rough surface is the surface roughness 
power spectrum $C({\bf q})$ defined by\cite{P3,Persson_JCP2001}
\begin{equation}
 \label{powerspectrum}
 C({\bf q}) = {1\over (2\pi )^2}
              \int d^2x \ \langle h({\bf x})h({\bf 0})\rangle 
              e^{i{\bf q}\cdot {\bf x}}.
\end{equation}
Here $\langle ... \rangle$ stands for ensemble average and we 
have assumed that $h({\bf x})$ is measured from the average surface plane so that 
$\langle h \rangle = 0$.
In what follows we will assume that
the statistical properties of the surface are isotropic, in which case $C(q)$ will only
depend on the magnitude $q=|{\bf q}|$ of the wavevector ${\bf q}$. 

\begin{figure}
\includegraphics[width=0.35\textwidth]{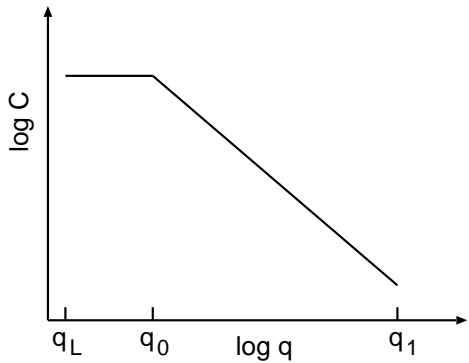} %{Cq.1.ps}
\caption{\label{Cq1}
Surface roughness power spectrum of a surface which is self-affine fractal for
$q_0<q<q_1$. The long-distance roll-off wavevector $q_0$ and the short distance cut-off
wavevector $q_1$ depend on the system under consideration. The slope of
the ${\rm log} C-{\rm log}q$ relation for $q > q_0$ determines the fractal
exponent of the surface. The lateral size $L$ of the surface (or of the studied surface region)
determines the smallest possible wavevector $q_L=2\pi /L$.}
\end{figure}

Many surfaces tend to be nearly self-affine fractal. A self-affine fractal
surface has the property that if part of the surface is magnified, with a magnification
which in general is appropriately different in the perpendicular direction to the surface as compared
to the lateral directions, then the surface ``looks the same'', i.e., the statistical
properties of the surface are invariant under this scale transformation\cite{P3}.
For a self-affine
surface the power spectrum has the power-law behavior
\[
 C(q) \sim q^{-2(H+1)},
\]
where the Hurst exponent $H$ is related to the fractal dimension $D_{\rm f}$ of the surface via
$H=3-D_{\rm f}$. Of course, for real surfaces this relation only holds in some finite
wavevector region $q_0 < q < q_1$, and in a typical case $C(q)$ has the form shown
in Fig.~\ref{Cq1}. Note that in many cases there is a roll-off wavevector $q_0$ below which
$C(q)$ is approximately constant. 

\begin{figure}
  \includegraphics[width=0.50\textwidth]{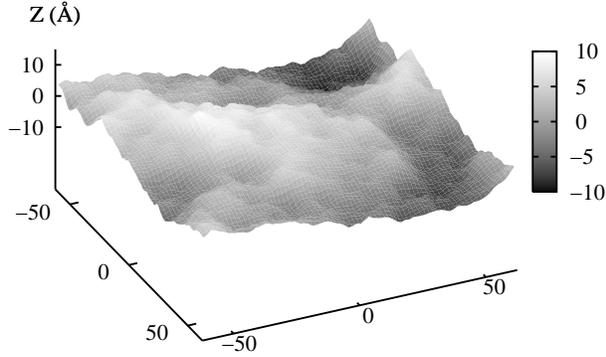} %{fractal_surface_01_fin1_1.eps}
  \caption{ \label{fractal_surface}
          Fractal surface of the substrate with the large cut-off wavevector 
          $q_1= 2\pi /b = 78\, q_0$. 
          For a square $124.8\,\mbox{\AA}\times 124.8\,\mbox{\AA}$ surface area.
          The fractal dimension $D_{\rm f} = 2.2$ and the root-mean-square roughness 
          amplitude is $3~\mbox{\AA}$.}
\end{figure}

In our calculations we have used self-affine fractal surfaces generated as outlined in 
Ref.~\cite{P3}. 
Thus, the surface height is written as
\begin{equation}
 \label{profile}
 h({\bf x}) =\sum_{\bf q} B({\bf q}) e^{i[{\bf q}\cdot {\bf x}+\phi({\bf q})]},
\end{equation}
where, since $h({\bf x})$ is real, $B(-{\bf q})=B({\bf q})$ and $\phi(-{\bf q})=
-\phi({\bf q})$. If $\phi({\bf q})$ are independent random variables, uniformly distributed
in the interval $[0,2\pi[$, then one can easily show that higher order
correlation functions involving $h({\bf x})$ can be decomposed into a product of pair correlations,
which implies that the height
probability distribution $P_h = \langle \delta (h-h({\bf x}))\rangle $ is Gaussian\cite{P3}.
However, such surfaces can have {\it arbitrary
surface roughness power spectrum}.
To prove this, 
substitute (\ref{profile}) into (\ref{powerspectrum}) and use that
$$\langle e^{i\phi({\bf q'})} e^{i\phi({\bf q''})}\rangle = \delta_{{\bf q'},-{\bf q''}}$$
gives
$$C({\bf q})={1\over (2\pi )^2} \int d^2x \ \sum_{\bf q'} |B({\bf q'})|^2 e^{i({\bf q}-{\bf q'})
\cdot {\bf x}}
$$
$$=
\sum_{{\bf q'}} |B({\bf q'})|^2 \delta ({\bf q}-{\bf q'}).$$
Replacing
$$\sum_{{\bf q}} \rightarrow {A_0\over (2\pi)^2}\int d^2q,$$
where $A_0$ is the nominal surface area, gives
$$C({\bf q})={A_0\over (2\pi)^2} |B({\bf q})|^2.$$
Thus, if we choose
\begin{equation}
 \label{Bq}
 B({\bf q})= (2\pi/L) [C({\bf q})]^{1/2},
\end{equation}
where $L=A_0^{1/2}$, then the surface roughness profile (\ref{profile}) has the
surface roughness power spectrum $C({\bf q})$. If we assume that the
statistical properties of the rough surface are isotropic, then $C({\bf q})=C(q)$
is a function of the magnitude $q=|{\bf q}|$, but not of the direction of ${\bf q}$.
The randomly rough substrate surfaces used in our numerical calculations were
generated using (\ref{profile}) and (\ref{Bq}) and assuming that the surface roughness
power spectra have the form shown in Fig.~\ref{Cq1},
with the fractal dimension $D_{\rm f}=2.2$ and 
the roll-off wavevector $q_0=3q_L$, where $q_L= 2\pi /L_x$.
We have chosen $q_0=3q_L$ rather than $q_0=q_L$ since the former value gives 
some self-averaging and less noisy numerical results.
We also used $q_1= 2\pi /b = 78q_0$. The topography of the substrate  
with the root-mean-square roughness amplitude $3~\mbox{\AA}$ used in our 
numerical calculations is shown in Fig.~\ref{fractal_surface}.

%\vskip 0.5cm
%{\bf 4. Numerical results}
\section{Numerical results}
\label{sec4}

In this section we present the results of molecular dynamics calculations of sliding
of elastic blocks on rigid substrates. In all cases, unless otherwise stated, 
the upper surface of the block 
moves with the velocity $v=0.1 \ {\rm m/s}$, and the (nominal) 
squeezing pressure $p$ is
one tenth of the elastic modulus $E$ of the block, i.e., $p=0.1E$. The reason for
choosing $p$ proportional to $E$ is twofold. First, we consider solids with elastic
modulus which varies over several orders of magnitude, and it is not possible to use a constant $p$
as this would result in unphysical large variations in the elastic deformation of the block.
Second, if two elastic solids are squeezed together with a given load, then as long as
the area of real contact is small compared to the nominal contact area, the 
pressure in the contact areas will be proportional to the elastic modulus of the
solids\cite{Persson_JCP2001}. 
Initially, when the block is pulled laterally it deforms loading elastic
energy and the shear force between the surfaces increases gradually. The
shear force reaches a maximum $F_s$ at the onset of sliding. We used to
such maximum shear force to calculate the static friction coefficient:
$\mu_s = F_s / F_N$, $F_N=pA$ being the normal force.
During the sliding the shear force oscillates in time. We defined the
kinetic friction coefficient as the ratio between the time averaged
shear force and the normal load, i.e., $\mu_k=\overline{F_k}/F_N$.

\begin{figure}
  \includegraphics[width=0.50\textwidth]{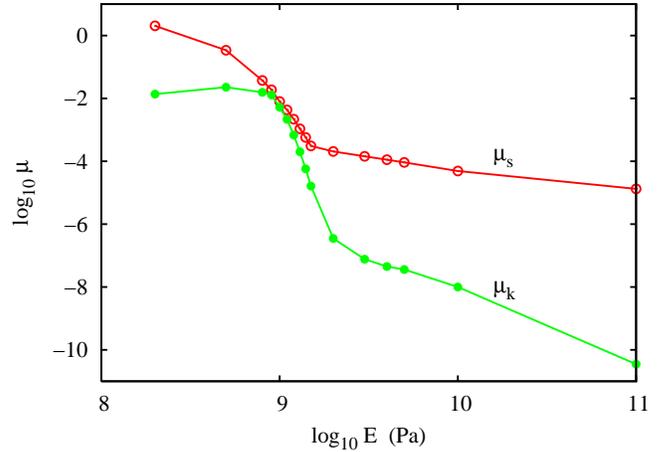} %{stat_and_kin_frict_coeff_Flat_with_and_without_adhesion_7_fin1_1.eps}
  \caption{ \label{stat_kin_Flat}
The static (curve 1) and kinetic (curve 2) friction coefficients as a function of the elastic modulus 
$E$ of the block, for the flat substrate. 
In the calculation we have assumed the squeezing pressure $\sigma_0 = 0.1 E$ and the
sliding velocity $v=0.1 \ {\rm m/s}$.
}
\end{figure}

Let us first assume that both the block and the substrate have atomically smooth surfaces.
Fig.~\ref{stat_kin_Flat} shows the static and the kinetic friction coefficients as a 
function of the elastic modulus 
$E$ of the block.
Note the relatively abrupt decrease in the friction when the elastic modulus changes
from $E_1\approx 0.7 \ {\rm GPa}$ to $E_2 \approx 2 \ {\rm GPa}$. 
For $E>E_2$ practically no instabilities occur and the
friction is extremely small,
while for $E < E_1$ relatively strong elastic
instabilities occur at the sliding interface, and the friction is high.
For $E=0.2 \ 
{\rm GPa}$ the static friction $\mu_{\rm s} > 2$. 
This calculation illustrates that the transition from high friction to {\it superlubricity} can 
be very abrupt; in the present case an increase in the elastic modulus by only a 
factor of $\sim 3$ (from 0.7 to $2.1 \ {\rm GPa}$) 
decreases the kinetic friction by a factor of $\sim 10^{5}$.

\begin{figure}
  \includegraphics[width=0.50\textwidth]{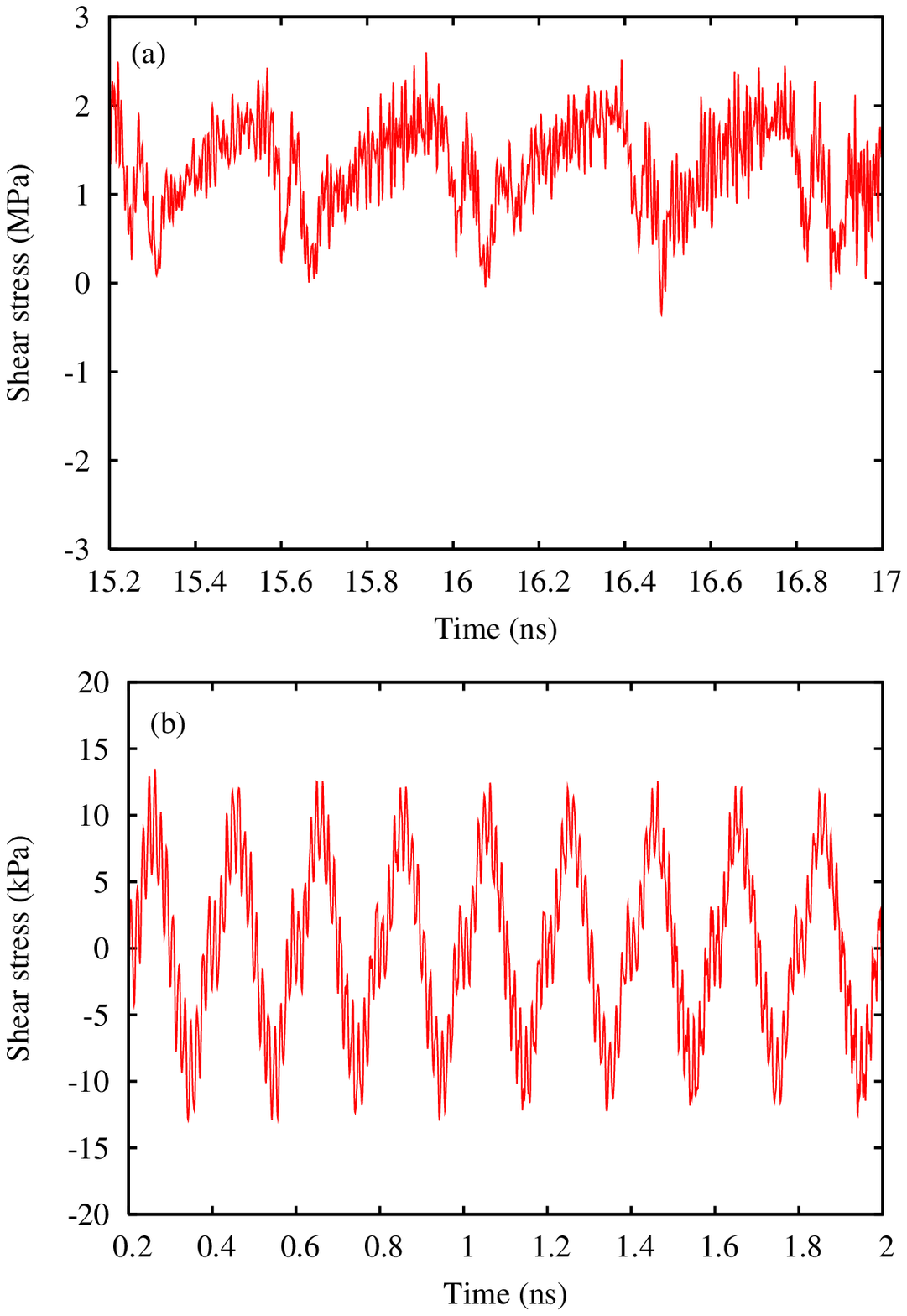} %{shear_E_0_8_2_0_GPa_fin_1_2.eps}
  \caption{ \label{shear.smooth1}
(a) The shear stress as a function of time for the flat substrate.  
The 
elastic modulus of the block is $E= 0.8 \ {\rm GPa}$. 
(b) The same as above but for 
the elastic modulus of the block $E= 2 \ {\rm GPa}$. 
}
\end{figure}

In Fig.~\ref{shear.smooth1} we show the time variation of
the shear stress as a function of time when the elastic modulus of the block
equals (a) $E= 0.8 \ {\rm GPa}$
and (b) $E= 2 \ {\rm GPa}$. 
The elastic modulus of the stiffer solid is above the superlubricity threshold, 
and no (or negligible) elastic instabilities occur;
the stress is a periodic function of time,
with the period corresponding to the displacement 
$0.2 \ \mbox{\AA}$. For the softer solid strong elastic instabilities 
occur during sliding, the shear stress
is less regular (and the arrangement of the interfacial block atoms more disordered)
than for the stiffer solid, and the (average) period is {\it longer}
than $0.2 \ \mbox{\AA}$.  

\begin{figure}
  \includegraphics[width=0.45\textwidth]{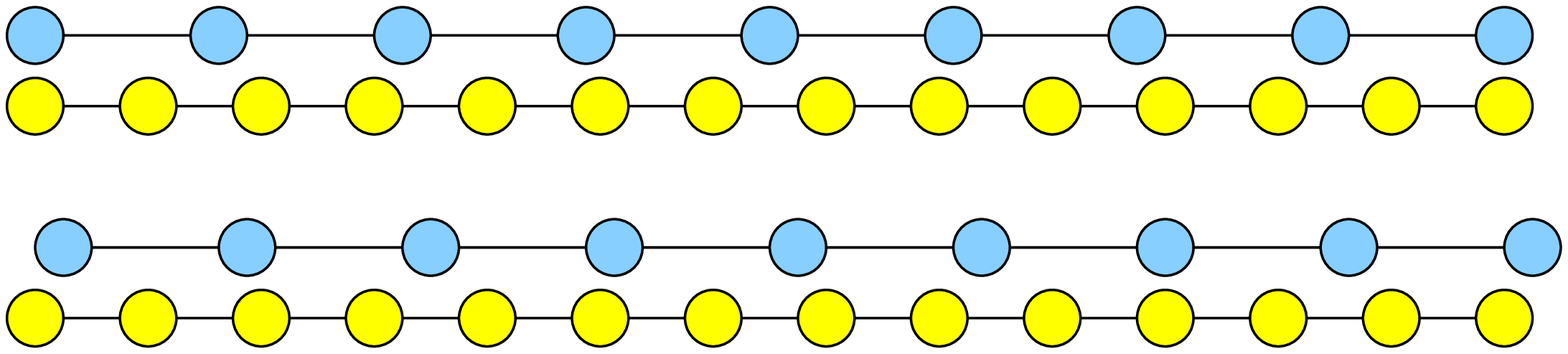} %{8_13.eps}
  \caption{ \label{config1}
Commensurability ratio $8/13$ between 1-dimensional chains.
The upper image shows the period of 8 block atoms (light blue), i.e.,
13 substrate atoms (yellow). The lower image is obtained by shifting
the block for $1/13$ of its lattice spacing. Block atoms occupy the same
positions relatively to the substrate's hollows.}
\end{figure}

The regular pattern with period 0.2 \AA\ in Fig.~\ref{shear.smooth1}(b)
can be understood as follows.
For our system, in the sliding direction 
there are 8 block atoms for every 13 substrate atoms. Assume first that the 
block (and the substrate) are perfectly stiff. In this case, the position of the 
8 block atoms
will
take 8 uniformly spaced positions within the substrate 
unit cell (lattice constant $b$), see Fig.~\ref{config1}. Thus,
a shift of the block with the distance $b/8$ will take the system to a (geometrically) equivalent
configuration. Hence, since $b=1.6 \ \mbox{\AA}$ 
we expect the periodicity of the shear stress to be $b/8   
=0.2 \ \mbox{\AA}$. When the block has a finite elasticity but above the
superlubricity threshold, the atoms will relax somewhat 
in the substrate potential, but the configuration of the system will still repeat itself
with the same period $b/8$. 

\begin{figure}
  \includegraphics[width=0.50\textwidth]{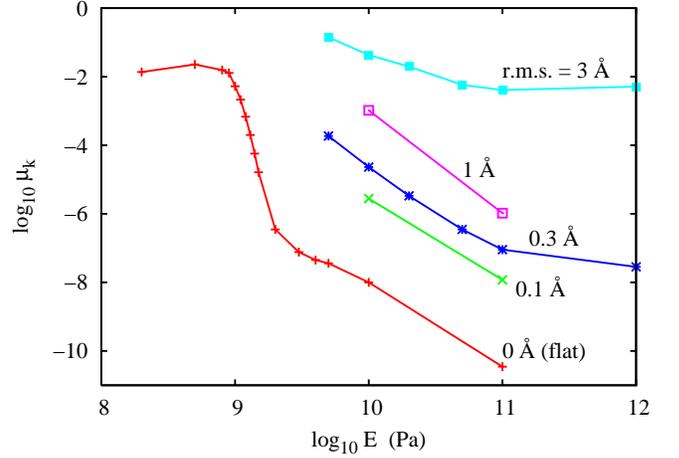} %{kin_frict_coeff_points_Flat_Fractal_vs_E_and_RMS_10_notitle_fin1_1.eps}
  \caption{ \label{KinFractal} 
The kinetic friction coefficients for an elastic block sliding on 
rough substrates, as a function of the 
logarithm of elastic modulus $E$ of the block. 
The curves 1-5 (from top to bottom) correspond to the root-mean-square roughness amplitudes of the 
fractal substrate $3$, $1$, $0.3$, $0.1~\mbox{\AA}$ and $0$ 
(flat substrate). 
}
\end{figure}

Let us now consider the influence of surface roughness on the sliding dynamics.
In Fig.~\ref{KinFractal} we show
the kinetic friction coefficients for an elastic block sliding on a
rough substrate, as a function of the 
logarithm of elastic modulus $E$ of the block. 
The curves from top to bottom correspond to the substrate root-mean-square roughness amplitudes 
$3$, $1$, $0.3$, $0.1~\mbox{\AA}$ and $0$ 
(flat substrate). For the substrate with the largest roughness, no superlubricity state can be observed
for any elastic modulus up to $E=10^{12} \ {\rm Pa}$.

\begin{figure}
  \includegraphics[width=0.50\textwidth]{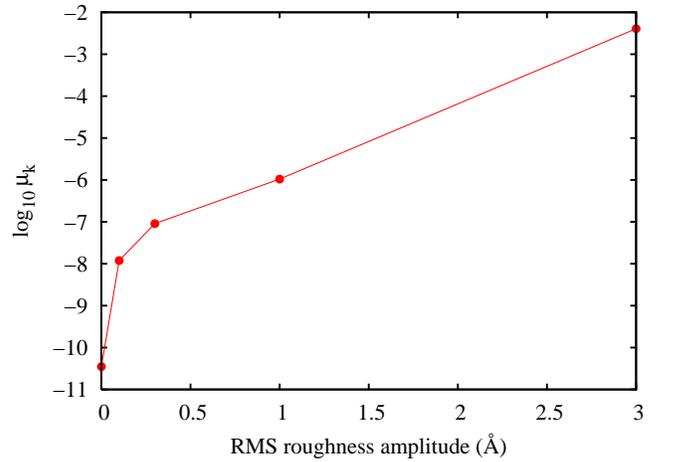} %{kin_frict_coeff_vs_RMS_s11b_Fractal_3_fin1_2.eps}
  \caption{ \label{kin1}
The kinetic friction coefficient as a function of the root-mean-square roughness amplitude of 
the substrate. 
The elastic modulus of the block $E=100 \ {\rm GPa}$. 
}
\end{figure}

In Fig.~\ref{kin1} we show
the kinetic friction coefficient as a function of the root-mean-square roughness amplitude of 
the substrate. 
The elastic modulus of the block $E=100 \ {\rm GPa}$. 
Note the strong decrease in the friction when the root-mean-square roughness amplitude decreases
below $0.3~\mbox{\AA}$, which corresponds to a peak-to-peak roughness of 
roughly one atomic lattice spacing.

\vskip 0.5cm

\begin{figure}
  \includegraphics[width=0.50\textwidth]{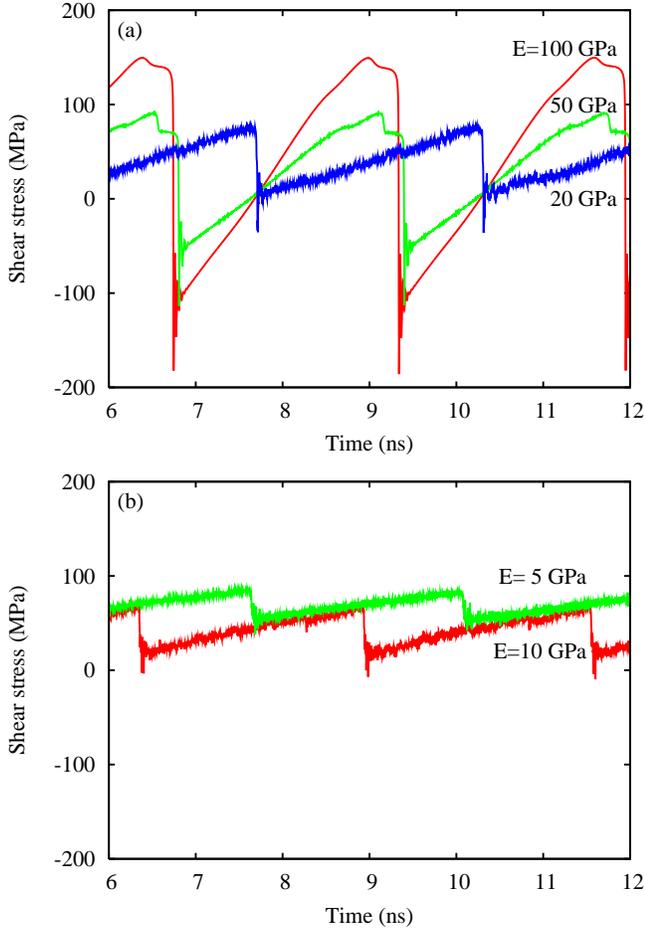} %{shear_vs_t_s11b_s10b5_s10b2_Fractal_fin1_1_fin.eps}
  \caption{ \label{shear1}
(a) The shear stress as a function of time for the rough substrate  
with root-mean-square roughness amplitude $3~\mbox{\AA}$. The 
elastic modulus of the block is $E=100$, $50$ and $20$ GPa. 
(b) The same as above but for 
the elastic modulus of the block $E=10$ and $5$ GPa. 
}
\end{figure}

Figure \ref{shear1}(a) shows the average (or nominal) shear stress
as a function of time for the rough substrate  
with the root-mean-square roughness $3~\mbox{\AA}$, and for the 
elastic modulus of the block $E=100 $, $50$ and $20 \ {\rm GPa}$. 
Note that in addition to major slip events, several small slip events occur in all 
cases. These events correspond to local slip at some asperity contact regions
before the major slip involving the whole contact area.
In all cases, the time dependence of the shear stress remains periodic with the period  
$2.6~\mbox{\AA}$, which corresponds 
to the lattice spacing of the block.
% To be compared with the spatial period 0.2 Angstrom for flat superlubric surfaces.
Note also that for the elastically softer
block ($E=20 \ {\rm GPa}$), the stress-noise increases 
after each major slip event; this is caused by the elastic waves
(heat motion) excited during the (major) rapid slip events and not
completely adsorbed by the thermostat. 

Fig.~\ref{shear1}(b) shows the same as in (a) but now for 
the elastic modulus of the block $E=10 $ and $5 \ {\rm GPa}$. 
In this case the decrease of the elastic modulus of the block results in the 
increase of both the static and kinetic friction. 

\begin{figure}
  \includegraphics[width=0.50\textwidth]{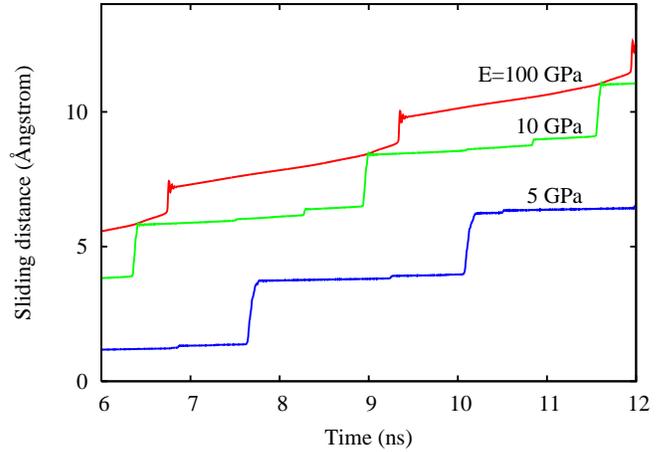} %{positions_1_14_s11b_s10b_s09b5_Fractal_fin1_1.eps}
  \caption{ \label{slipE}
The average displacement of the interfacial atoms of the block  
as a function of time.  The root-mean-square roughness amplitude for the
substrate is $3~\mbox{\AA}$. The 
elastic modulus of the block is $E=100$, $10$ and $5$ GPa. 
}
\end{figure}

Fig.~\ref{slipE} shows the average displacement of the interfacial atoms of the block 
(in the sliding direction) as a function of 
time.
The 
root-mean-square roughness amplitude for the substrate is $3~\mbox{\AA}$. The 
elastic modulus of the block is $E=100$, $10$ and $5 \ {\rm GPa}$. 
Note that the slip distance for the major slip events increases as 
the elastic modulus of the block decreases,
and that for the elastically hardest solid ($E=10^{11} \ {\rm Pa}$) about a half of the forward displacement occurs 
between the major slip events.

\begin{figure}
  \includegraphics[width=0.50\textwidth]{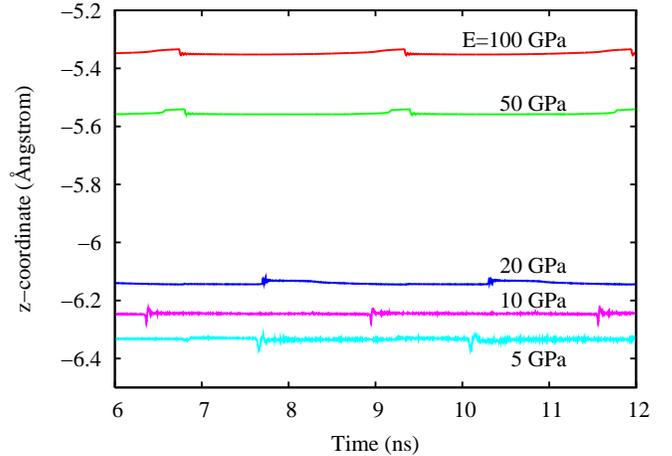} %{positions_1_16_s11b_s10b5_s10b2_s10b_s09b5_Fractal_paper_fin1_1.eps}
  \caption{ \label{height1}
The average position of the interface block atoms in the $z$-direction (perpendicular to the 
sliding direction), as a function of time during sliding. 
The root-mean-square roughness amplitude of the substrate is $3~\mbox{\AA}$.
The curves from top to bottom correspond to 
the elastic modulus of the block $E=100$, $50$, $20$, $10$ and $5$ GPa.
}
\end{figure}

Fig.~\ref{height1}
shows the average position of the interface block atoms in the $z$-direction (perpendicular to the 
sliding direction) as a function of time.
Results are shown for the rough substrate with the root-mean-square roughness amplitude 
$3~\mbox{\AA}$. The elastic modulus of the block is $E=100$, 
$50$, $20$, $10$ and $5 \ {\rm GPa}$. When the elastic
modulus decreases, because of the adhesive interaction the 
block interfacial atoms come (on the average) more close to the {\it 
rigid} substrate, embracing the substrate asperities. This increases  the 
real area of contact between the surfaces and results in a higher friction. 

\begin{figure}
  \includegraphics[width=0.50\textwidth]{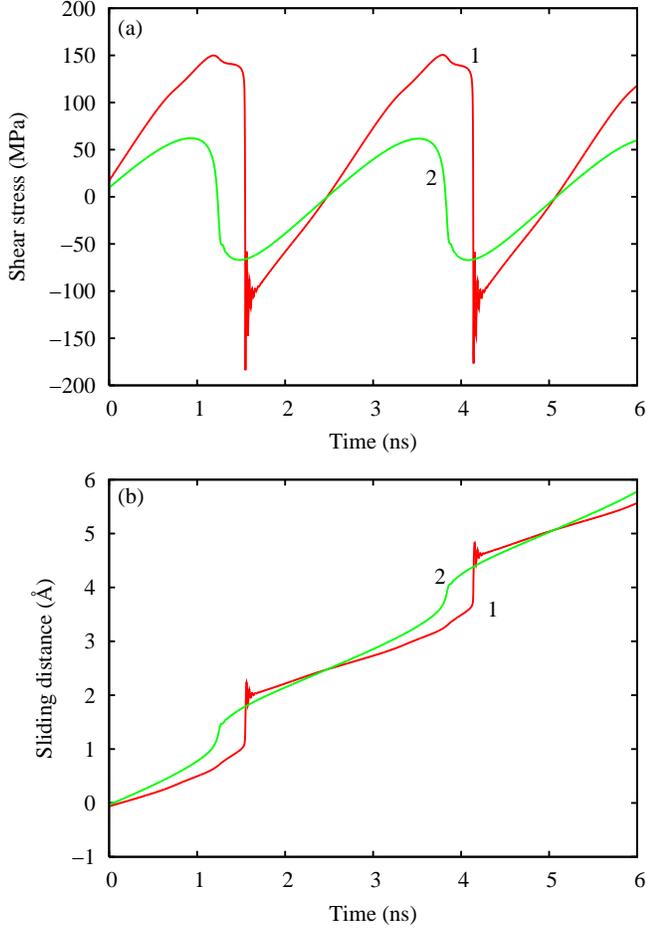} %{shear_vs_t_s11b_s11b_without_adhesion_Fractal_fin1_1_fin.eps}
  \caption{ \label{shear6}
(a) The shear stress as a function of time calculated for the rough substrate 
(root-mean-square amplitude $3 \ \mbox{\AA}$) and for the stiff block ($E = 100 \ {\rm GPa}$). 
Curve 1 is calculated including adhesion, while Curve 2 is obtained without the attractive part
in the Lennard-Jones potential, i.e., with $\alpha = 0$ in Eq.~(\protect\ref{potential}). 
(b) The average displacement of the interface block atoms (in the sliding direction) as
a function of 
time for the same systems as in (a).
}
\end{figure}

Fig.~\ref{shear6}(a) shows the shear stress as a function of time for the rough substrate 
(root-mean-square amplitude $3 \ \mbox{\AA}$) and for the stiff block ($E = 100 \ {\rm GPa}$). 
The solid curve is with the adhesion included, while the dashed curve is without the attractive part
in the Lennard-Jones potential, i.e., with $\alpha = 0$ in Eq.~(\ref{potential}). Note that
without adhesion the major slip is not so pronounced as for the case with 
adhesion. Still the time dependence of the shear stress remains periodic with the same 
period $2.6~\mbox{\AA}$, 
corresponding to the lattice spacing of the block. Without adhesion, the shear stress curve is nearly symmetric
around the zero-stress axis, and the kinetic friction coefficient (determined by the average shear stress
divided by the squeezing pressure) is about 150 times smaller than when the adhesive
interaction is included. 

In Fig.~\ref{shear6}(b) we show 
the average displacement of the interface block atoms (in the sliding direction) as
a function of 
time for the same systems as in (a).
For the case without adhesion the major slip is 
not as abrupt as when adhesion is included. At every moment there is some lateral motion 
of the block interfacial atoms.

\begin{figure}
  \includegraphics[width=0.50\textwidth]{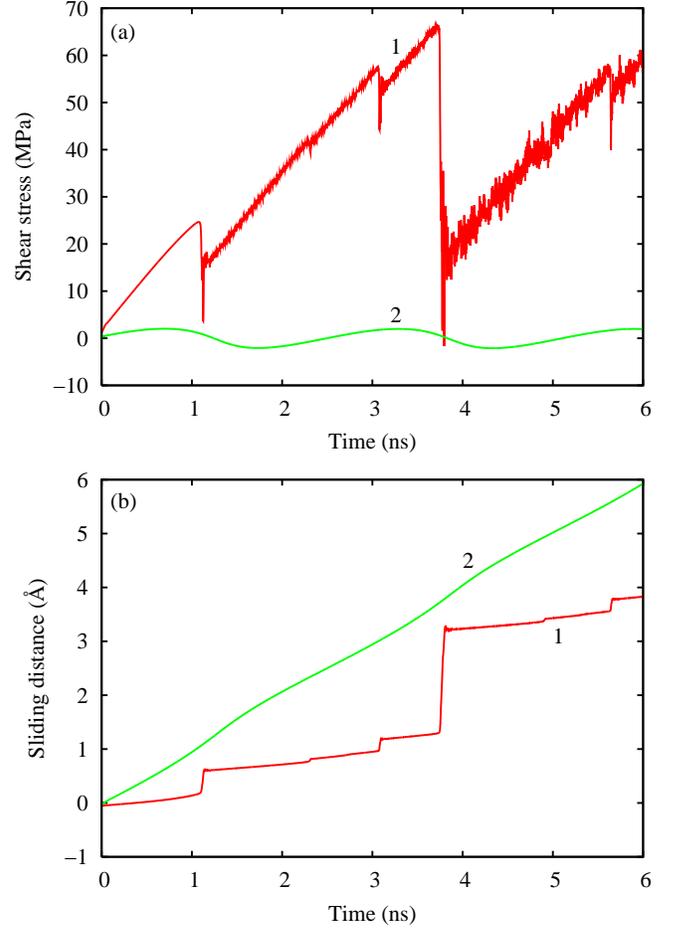} %{shear_vs_t_s10b_s10b_without_adhesion_Fractal_fin1_1_fin.eps}
  \caption{ \label{shear9}
(a) The shear stress as a function of time for the rough substrate 
(root-mean-square amplitude $3 \ \mbox{\AA}$) and for the block elastic modulus $E = 10 \ {\rm GPa}$. 
Curve 1 is calculated including adhesion, while Curve 2 is obtained without the attractive part
in the Lennard-Jones potential, i.e., with $\alpha = 0$ in Eq.~(\protect\ref{potential}). 
(b) The average displacement of the interface block atoms (in the sliding direction) as
a function of 
time for the same systems as in (a).
}
\end{figure}

Fig.~\ref{shear9}(a) shows the shear stress as a function of 
time for the rough substrate 
(root-mean-square amplitude $3 \ \mbox{\AA}$) and for the elastic block with the
elastic modulus $E = 10 \ {\rm GPa}$. 
The solid curve is with adhesion included, while the dashed curve is without the attractive part
in the Lennard-Jones potential, i.e., with $\alpha = 0$ in Eq.~(\ref{potential}). 
Fig.~\ref{shear9}(b) shows the average displacement of the 
interface block atoms (in the sliding direction) as
a function of 
time for the same systems as in (a).
For the case without adhesion the major slip is 
not as abrupt as for the case with adhesion,
and the sliding motion is nearly steady. 
In both cases, the 
time dependence of shear stress remains periodic with the period 
$2.6~\mbox{\AA}$ determined by the lattice 
spacing of the block. 
For the case with adhesion two small slips and a 
major slip can be observed in each period, and the 
kinetic 
friction is high. For the case without adhesion 
no elastic instability occurs, and the kinetic friction is very small.

The roughness-induced 
increase of friction can be understood by considering that the real
contact involves only small regions, as shown in Fig.~\ref{contact1}.
The compensation of the lateral forces that guarantees superlubricity
between incommensurate surfaces (see Fig.~\ref{instabil}) does not
happen at the boundaries of the contact regions, neither it can happen
on very small contacts with high curvature.
Thus, the friction force should increase with increasing 
length of the boundaries between the contact and non-contact regions.
This is completely different from what happens between commensurate
walls or between walls with very strong interactions, e.g., metals
with cold-welded microjuctions, where the real area of
contacts determines the friction force.

The surfaces used in our simulations are self-affine fractal up
to the atomic scale, but in general the cut-off wavevector $q_1$ can be
smaller, so that the typical size of the contact areas can be larger and
the friction can be lower. In other words, the root-mean-square surface
roughness alone is not enough to determine the amount of friction: a
surface profile with wider mountains and valleys has to provide less
friction than a surface profile dominated by short wavelength
corrugation. In the same way, a surface with higher fractal dimension
will have more roughness at the smallest wavelength, providing higher
friction for the same root-mean-square roughness and cut-off wavevector
$q_1$. In particular for high fractal dimensions the friction must
depend dramatically on the cut-off wavevector.

%\vskip 0.5cm
%{\bf 5. Pressure dependence of the frictional stress}
\section{Pressure dependence of the frictional stress}
\label{sec5}

\begin{figure}
  \includegraphics[width=0.50\textwidth]{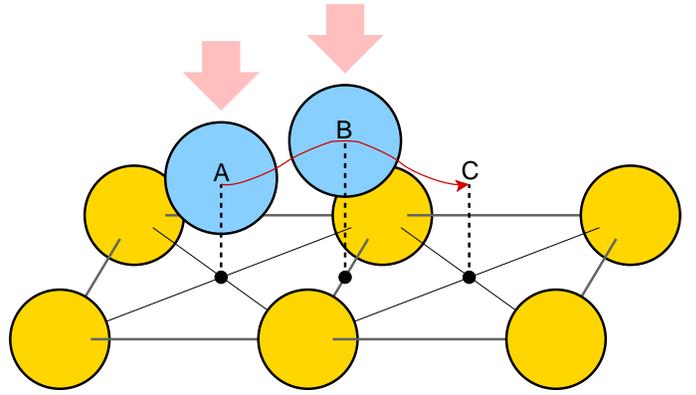} %{jump.eps}
  \caption{ \label{potential1}
A block atom moving (or jumping) from the hollow site {\bf A} over the bridge site {\bf B}
to the hollow site {\bf C}. The maximum energy position along the trajectory is at site {\bf B}.
}
\end{figure}

During sliding, the atoms at the sliding interface will experience energetic barriers
derived from both the adhesive interaction between the atoms 
on the two opposing surfaces,
and from the applied load. Thus, we may define an {\it adhesion pressure} 
$p_{\rm ad}$, and as long as
$p_{\rm ad} \gg p$, where $p$ is the pressure in the contact area derived from
the external load, the frictional shear stress will be nearly independent of the applied
load. Let us illustrate this with the system studied in Sec.~\ref{sec4}. Let us first consider
the limiting case where the elastic modulus of the block is extremely small. In this
case, in the initial pinned state (before sliding) all the block atoms will occupy
hollow sites on the substrate, as indicated by atom {\bf A} in Fig.~\ref{potential1}. During
sliding along the $x$-direction, the atom {\bf A} will move over the bridge
position {\bf B} and then ``fall down'' into the hollow position {\bf C} (we assume overdamped
motion). The minimum energy for this process is given by the barrier height $\delta \epsilon$
(the energy difference between the sites {\bf B} and {\bf A}) plus the work $pa^2 \delta h$
against the external load, where $a$ is the block lattice constant and $\delta h$ the change in the
height between sites {\bf B} and {\bf A} (which depends on $p$). Thus the frictional
shear stress $\sigma_{\rm f}$ is determined by $\sigma_{\rm f} a^2 b = \delta \epsilon+pa^2\delta h$, or 
$$\sigma_{\rm f} = \delta \epsilon /(ba^2) +p\delta h/b = (p_{\rm ad}+p)\delta h /b,$$
where we have defined the adhesion pressure $p_{\rm ad} = \delta \epsilon/(a^2 \delta h)$.

In our case $\delta \epsilon \approx 3 \ {\rm meV}$ and $\delta h \approx 0.008 \ \mbox{\AA}$
giving $p_{\rm ad} \approx 10^{10} \ {\rm Pa}$. Thus, in the present case,
only when the local pressure in the
contact regions becomes of the order of $\sim 10 {\rm GPa}$, or more, it will start to 
influence the shear stress. This result is in accordance with our simulation results.
Thus, for smooth surfaces, the shear stress acting on the block with the elastic modulus
$E=0.5 \ {\rm GPa}$, squeezed against the substrate with the pressure $p=50$ and
$150 \ {\rm MPa}$, is identical 
($\approx 1 \ {\rm MPa}$) within the accuracy of the simulations.

For inert materials such as rubber
the adhesive pressure may be of similar magnitude as
obtained above. Since the contact pressure for rubber in most cases is below
$10 \ {\rm MPa}$, one may expect that 
the shear stress in the areas of real contact will
be independent on the load. Recently, a strong dependence of the (apparent) shear stress on
the squeezing pressure was observed for smooth Plexiglas balls sliding on very smooth
silicon wafers covered by silane layers\cite{Bureau}. However, as 
one of us has argued elsewhere\cite{P05},
this does not reflect a fundamental dependence of the shear stress on the squeezing pressure,
but has another origin. 

%\vskip 0.5cm
%{\bf 6. Summary and conclusion}
\section{Summary and conclusion}
\label{sec6}

We have studied the sliding of elastic solids in adhesive contact with flat and
rough interfaces. We considered the dependence of the
sliding friction on the elastic modulus of the solids.
For elastically hard solids with planar surfaces with incommensurate surface
structures we observe extremely low friction (superlubricity), which
very abruptly increases as the elastic modulus decreases. 
Thus, at the superlubricity threshold,
an increase in the elastic modulus by a factor of $\sim 3$ resulted in
the decrease in the frictional shear stress by a factor $\sim 10^{5}$.
We have shown that
even a relatively small surface roughness may completely 
kill the superlubricity. For flat surfaces the shear stress is independent
of the perpendicular (squeezing) pressure as long as the pressure $p$ is below the
adhesive pressure $p_{\rm ad}$, which typically is of the order of several GPa.

%\vskip 0.5cm

%{\bf Acknowledgments}
\section*{Acknowledgments}

A part of the present work was carried out in frames of the ESF 
program ''Nanotribology (NATRIBO)''. Two of the authors (U.T. and 
V.N.S.) acknowledge support from IFF, FZ-J\"ulich, hospitality 
and help of the staff during their research visits. The authors 
also thank V.V. Samoilov for technical assistance with preparing 
some figures for the paper. 
This work was partly sponsored by MIUR FIRB RBAU017S8 R004, MIUR FIRB
RBAU01LX5H, MIUR COFIN 2003 and PRIN-COFIN 2004.


\vspace{1em}

\begin{thebibliography}{999}

\bibitem{P0}
B.N.J. Persson,
{\it Sliding Friction: Physical Principles and Applications},
2nd ed., Springer, Heidelberg, 2000.

\bibitem{P11}
B.N.J. Persson, O. Albohr, F. Mancosu, V. Peveri, V.N. Samoilov and I.M. Sivebaek,
Wear {\bf 254}, 835 (2003).

\bibitem{P2}
B.N.J. Persson, 
Phys. Rev. B{\bf 51}, 13568 (1995).

\bibitem{Caroli}
C. Caroli and P. Nozieres, Eur. Phys. J. B{\bf 4}, 233 (1998).

\bibitem{Caroli1}
T. Baumberger and C. Caroli,
arXiv:cond-mat/0506657 v1 (2005). 

\bibitem{Aubry}
J. Aubry, 
J. Phys. (Paris) {\bf 44}, 147 (1983).

\bibitem{Japan}
K. Shinjo and M. Hirano, Surf. Sci. {\bf 283}, 473 (1993).

\bibitem{Elisa}
E. Riedo and H. Brune,
Applied Physics Letters {\bf 83}, 1986 (2003).

\bibitem{Flipse}
R.J.A. van der Oetelaar and C.F.J. Flipse, Surf. Sci. {\bf 384}, L828 (1997).

\bibitem{SSC}
B.N.J. Persson and E. Tosatti,
Solid State Communications {\bf 109}, 739 (1999).

\bibitem{Nozier}
C. Caroli and P. Nozieres, in {\it Physics of Sliding Friction},
ed. by B.N.J. Persson and E. Tosatti, Kluwer, Dordrecht (1996).

\bibitem{Muser}
M.H. M\"user, Europhys. Lett {\bf 66}, 97 (2004).

\bibitem{Meyer1}
E. Gnecco, R. Bennewitz, T. Gyalog, Ch. Loppacher, M. Bammerlin, E. Meyer and H.-J. G\"untherodt,
Phys. Rev. Lett. {\bf 84}, 1172 (2000); 
E. Riedo, E. Gnecco, R. Bennewitz, E. Meyer and H. Brune,
Phys. Rev. Lett. {\bf 91}, 084502 (2003). 

\bibitem{Activated}
Y. Sang, M. Dube and M. Grant,
Phys. Rev. Lett. {\bf 87}, 174301 (2001).

\bibitem{he1999}
G. He, M.H. M\"user and M.O. Robbins, Science {\bf 284}, 1650 (1999).
% Adsorbed Layers and the Origin of Static Friction
% http://dx.doi.org/10.1126/science.284.5420.1650

\bibitem{Frenken}
M. Dienwiebel, G.S. Verhoeven, N. Pradeep, J.W.M. Frenken, J.A. Heimberg
and H.W. Zandbergen, Phys. Rev. Lett. {\bf 92}, 126101 (2004).

\bibitem{Liu}
Y. Liu, A. Erdemir and E.I. Meletis, Surf. Coat. Technol. {\bf 86-87}, 564 (1996).

\bibitem{Erdemir}
The properties of diamond like carbon (DLC) films depend strongly on the preparation method and
operation conditions. Thus, only DLC films produced from discharge plasmas containing much hydrogen
will exhibit a low friction ($\mu \sim 0.001-0.003$). This is believed to result from the passivation of
carbon dangling bonds by the hydrogen atoms. Without hydrogen, in an inert atmosphere
the friction is huge (of the order of 1) 
because of a high concentration of very reactive carbon dangling bonds. In the normal atmosphere,
most dangling bonds are passivated and the friction lower but still much higher than for
diamond or for DLC films produced from plasmas containing much hydrogen.
See, A. Erdemir, Surface and Coatings Technology {\bf 146-147}, 292 (2001).

\bibitem{Yang}
C. Yang, U. Tartaglino and B.N.J. Persson,
Eur. Phys. J. E {\bf 19}, 47 (2006).

\bibitem{cai}
W. Cai, M. de Koning, V.V. Bulatov and S. Yip,
Phys. Rev. Lett. {\bf 85}, 3213 (2000).

\bibitem{E}
W. E and Z. Huang,
Phys. Rev. Lett. {\bf 87}, 135501 (2001).

\bibitem{Persson_JCP2001} 
B.N.J. Persson, J. Chem. Phys. {\bf 115}, 3840 (2001).

\bibitem{Jon}
K.L. Johnson, {\it Contact Mechanics} (Cambridge University Press,
Cambridge, 1985).

\bibitem{Is}
J. Israelachvili, {\it Intermolecular and Surface Forces} (Academic Press, London, 1992).

\bibitem{P3}
B.N.J. Persson, O. Albohr, U. Tartaglino, A.I. Volokitin and E. Tosatti, 
J. Phys. Condens. Matter {\bf 17}, R1 (2005). 

\bibitem{Bureau}
L. Bureau, T. Baumberger and C. Caroli,
Eur. Phys. J. E {\bf 19}, 163 (2006)
% DOI: 10.1140/epje/e2006-00019-2
% arXiv:cond-mat/0510232 (2005).

\bibitem{P05}
B.N.J. Persson, Surf. Sci. Reports, in press; ArXiv:cond-mat/0603807 (2006)

\end{thebibliography}
\end{document}